# THE INFLUENCE OF THE ACADEMY OF GONDISHAPUR ON THE FORMATION OF THE HOUSE OF WISDOM IN BAGHDAD AND THE SCIENCE OF THE GOLDEN AGE OF ISLAM (8th–10th CENTURIES)

**Rizoi Bakhromzod**
S. U. Umarov Physical-Technical Institute of the National Academy of Sciences of Tajikistan
Institute of Astrophysics of the National Academy of Sciences of Tajikistan
E-mail: rizo@physics.msu.ru
ORCID ID: 0000-0001-8894-8274

**Abstract**: The purpose of this study is to determine the role of the Academy of Gondishapur in the formation of the Bayt al-Hikma (House of Wisdom) in Baghdad and the emergence of Islamic science in the 8th–10th centuries. The methodology combines comparative-historical and biographical analysis with a study of institutional continuity based on primary sources and contemporary scholarly literature. Results demonstrate that Gondishapur – the foremost scientific and medical centre of late Sasanian Iran – constituted the primary human, methodological and institutional source for the Abbasid Bayt al-Hikma. The migration of specialists, especially the Bukhtishu dynasty, Yuhanna ibn Masawayh and Hunayn ibn Ishaq, transferred Gondishapur's medical traditions, hospital organisation (bimaristans) and translation methodology to Baghdad. Gondishapur's experience also shaped the development of medical education and the compilation of pharmacopoeias within the Bayt al-Hikma. A comparative structural analysis of the two institutions reveals both institutional borrowings and the original features of the Baghdad model. The article concludes that the transmission of Gondishapur's heritage to the Bayt al-Hikma constituted sustained institutional continuity, laying the foundations for the large-scale synthesis of Greek, Persian and Indian knowledge in the Arabic language.

## INTRODUCTION

The history of science in the Islamic Golden Age has long attracted researchers, yet the mechanisms of knowledge transmission from the pre-Islamic world to Abbasid Baghdad remain a subject of debate. The Academy of Gondishapur — the outstanding scientific-medical centre of late Sasanian Iran — is recognized as the key intermediary in this process [1, 2, 3, 4]. Founded presumably in the 3rd century CE under Shahanshah Shapur I and significantly expanded under Shapur II, this academy reached its peak during the era of Khosrow I Anushirvan (531–579 CE) [5, 6]. Here, a hospital, a school of higher education, and an extensive library functioned simultaneously; here scholars of diverse peoples and faiths — Zoroastrians, Nestorians, Jews, Greeks and Indians — laboured together [19, 20].

After the Arab conquest of Iran (651 CE), Gondishapur continued to function as the leading medical centre; however, by the mid-8th century intellectual leadership began to pass to the new Abbasid capital — Baghdad [7, 8]. In 762, Caliph al-Mansur founded Baghdad, and later his successors — Harun al-Rashid and especially al-Mamun (813–833) — established at the court the Bayt al-Hikma ("House of Wisdom"), which became the era's foremost centre of translation and scholarly activity [1, 9, 10]. The relevance of studying the continuity between the two institutions is determined not only by their historical significance, but also by the methodological questions concerning the nature of institutional knowledge transmission at the intersection of civilisations [11].

The purpose of this article is a comprehensive study of the influence of the Academy of Gondishapur on the formation of Bayt al-Hikma in Baghdad. To achieve this goal, the following tasks have been set: (1) to characterise the Academy of Gondishapur as an institution and determine the causes of its decline; (2) to trace the migration of scholars and the mechanisms of personnel continuity; (3) to analyse the translation movement as a channel for the transmission of knowledge; (4) to reveal the contribution of specific scholars of Iranian and Central Asian origin; (5) to conduct a comparative structural analysis of both institutions.

## MATERIALS AND METHODS

The source base of the study consists of classical Arabic-language bio-bibliographical compilations — primarily the "Fihrist" of Ibn al-Nadim and the "ʿUyun al-anba fi tabaqat al-atibba'" of Ibn Abi Usaybi'a — as well as hagiographic works preserving information about translators and court physicians of the Abbasid era. The secondary base includes fundamental monographs on the history of Islamic medicine [12, 13, 4], the history of the translation movement [1, 11, 14], and the history of Arabic mathematics [15, 16, 17].

The principal methodological tool is comparative-historical analysis, which allows comparison of the structure and functions of Gondishapur and Bayt al-Hikma. The biographical method (for reconstructing the migration routes of scholars and the mechanisms of knowledge transmission) and the method of institutional analysis (for identifying borrowed and original elements of the Baghdad model) are also applied. The interpretation of material is constructed with due consideration of critical revisions of traditional narratives in the works of D. Gutas [1], M. Dols [18], and A.I. Sabra [11], who caution against a simplistic understanding of "linear" knowledge transmission.

## RESULTS

1. The Academy of Gondishapur: Flourishing and Historical Context

The Academy of Gondishapur, situated in the province of Khuzestan in southwestern Iran, was a unique institution for late Antiquity, combining the features of a university, a research centre, and a clinical hospital [19, 20]. According to historians and contemporary studies, under Shahanshah Khosrow I the academy may have had several hundred teachers and thousands of students [21, 22]. A particularly significant role in the institution's flourishing was played by the fact that after the closure of the Platonic Academy in Athens (529 CE) and the persecution of Nestorian scholars in Byzantium, some Greek philosophers and Christian scholars relocated to Iran [23].

The academy comprised: a school of higher education with a multinational academic staff; a bimaristan (hospital), which served as the clinical base for training physicians directly at the bedside; an extensive library with a rich manuscript collection; and, presumably, an observatory. The hospital at Gondishapur is considered one of the first teaching hospitals in history, where treatment and systematic clinical training were combined [24, 22, 25]. D. Miller emphasises that the Gondishapur model of "a teaching hospital attached to a school" was revolutionary for the late Antiquity period: unlike European monastic hospices, it presupposed the systematic training of medical specialists [24].

The multi-confessional and multi-ethnic composition of Gondishapur's scholars predetermined the character of knowledge transmission: Greek philosophy and medicine, Indian astronomy and mathematics, Iranian and Mesopotamian traditions were synthesised here into a unified academic culture [2, 3]. A particular role in replenishing Gondishapur's library was played by Sanskrit texts: the court physician Borzuy, commissioned by Khosrow I, undertook an expedition to India and brought back medical and philosophical works, translating, in particular, the "Panchatantra" into Pahlavi [5, 26, 27]. Thus Sasanian Iran became an intermediary between the ancient and Eastern intellectual heritage, on the one hand, and the nascent Islamic science, on the other [28, 2].

In 636 CE, during the Arab conquest, Gondishapur came under the authority of the new caliphate. Despite the political upheaval, the academy did not cease its activities: the Arab rulers quickly recognized the value of the knowledge and personnel accumulated there [7]. According to the "Cambridge History of Iran", the school and hospital of Gondishapur "played an enormous role in the transmission of Greek, Iranian and Indian scientific knowledge to the Islamic world" [6, 7]. Nevertheless, by the end of the 8th — beginning of the 9th century, the scientific centre of gravity of the caliphate shifted steadily to Baghdad, and Gondishapur began to decline [8].

2. The Migration of Scholars: The Bukhtishu Family and Others

The personnel transmission of knowledge from Gondishapur to Baghdad was accomplished primarily through the targeted recruitment of Persian scholars to the courts of Abbasid caliphs. The most vivid example of this process is the dynasty of Perso-Nestorian physicians Bukhtishu ("servant of Jesus"), spanning six generations (7th–9th centuries) [29, 30]. Its members initially worked in Gondishapur, were versed in Greek and Indian medical science, and participated in translating classical works into Syriac and Pahlavi [12, 13].

The founder of the Baghdad branch of the dynasty — Jurjis (George) ibn Bukhtishu, director of the Gondishapur hospital — was summoned by Caliph al-Mansur around 765 CE to treat a severe illness [31, 32]. Having successfully fulfilled his appointment, he initiated a "brain drain" from Gondishapur to the capital. His son — Bukhtishu II — became the court physician of Harun al-Rashid, and his grandson — Jibra'il ibn Bukhtishu — around 800 CE founded in Baghdad the first large bimaristan on the instructions of that same caliph [18, 33, 29]. Thus, not only medical practice but also the organisational model of the Gondishapur hospital were transplanted to new soil.

In parallel with the Bukhtishu family, Yuhanna ibn Masawayh (c. 777–857 CE), a native and product of the Gondishapur school, moved to Baghdad. His father Masawayh was a pharmacist and teacher at the academy, which conditioned Yuhanna's deep immersion in its traditions [34, 35]. Subsequently, Masawayh became the court physician of several Abbasid caliphs, headed the palace hospital, and compiled in Arabic the first systematic treatise on ophthalmology, tracing back to the Gondishapur school [34, 36, 37].

3. Bayt al-Hikma: Structure and Functions

Bayt al-Hikma initially arose as a palace library ("khazanat al-hikma") under Harun al-Rashid, but under al-Mamun (813–833 CE) acquired the status of a full-fledged academy and translation centre [1, 9]. D. Gutas cautions against a simplistic understanding of this institution: Bayt al-Hikma was not a "translation academy" in the modern sense, but to a considerable extent functioned as a book repository at the caliph's court [1]. Nevertheless, it was here that the conditions for systematic translation activity were established, which defined the character of the "Islamic Golden Age" [14].

The institution was headed by a scholar-administrator associated with the caliphal court; subordinate to him worked translation teams specialising by field. The participants of the translation project belonged to different cultural and religious communities — Nestorian Christians, Sabians, Zoroastrians, Jews and Muslims [1, 38, 39, 40]. This reproduced the multi-confessional spirit of Gondishapur. The scale of activity was impressive: according to contemporary studies, Bayt al-Hikma employed dozens of translators working with Greek, Syriac, Persian and Sanskrit texts [1, 38].

4. The Translation Movement: Methodology and Key Texts

The translation movement of the 8th–10th centuries represents one of the most grandiose knowledge transfers in human history. A.I. Sabra characterised it as the "appropriation and naturalisation" of Greek science by Islamic civilisation: the acquired knowledge was not simply reproduced, but creatively reworked and adapted [11]. D. Gutas showed that the translation project had quite specific political and ideological motives — the Abbasid caliphs sought to assert their superiority over Byzantium and to draw upon the heritage of Iranian kings [1].

The infrastructure of translation activity traced back to Gondishapur: it was there that Nestorian scholars, as early as the 6th–7th centuries, had begun translating Greek medical texts into Syriac [23, 39, 40]. The method was twostage — first from Greek into Syriac (or from Sanskrit into Pahlavi), then into Arabic [39]. This practice was reproduced in Baghdad. The subject matter of translations encompassed medicine (Galen, Hippocrates, Dioscorides), philosophy (Aristotle, Plato), mathematics (Euclid), astronomy (Ptolemy) and Indian astronomical works [41, 28, 42].

The central figure of this process was Hunayn ibn Ishaq (809–873 CE) — a Syrian Nestorian Christian who became the foremost translator of the 9th century. A student of Yuhanna ibn Masawayh (himself a product of Gondishapur), Hunayn personally translated more than a hundred scholarly works, including an extensive corpus of Galen's writings — from Greek into Syriac and Arabic [43, 44, 45, 46]. He developed a rigorous philological method: comparing several manuscripts, verifying translations, and working with terminology [47, 48, 49, 50]. It was precisely this method, tracing back to the Nestorian tradition of Gondishapur, that made possible the high quality of Arabic translations of Greek medicine [51].

Indian astronomical traditions penetrated Islamic science also through a channel connected with Gondishapur. D. Pingree showed that Sasanian Iran was the key intermediary in the transmission of Indian astronomy to the West [26, 27]. The Sanskrit astronomical treatise "Brahmasphutasiddhanta" was translated into Arabic around 770 CE under Caliph al-Mansur, giving impetus to the flourishing of mathematical astronomy in Baghdad [52].

5. Iranian and Central Asian Scholars in Bayt al-Hikma

One of the distinguishing features of the science of the Abbasid period was the dominance of scholars of nonArab — predominantly Tajik (Iranian) — origin [2, 14]. This reflected the caliphs' deliberate policy of attracting the "best minds" regardless of ethnic background.

Muhammad ibn Musa al-Khwarizmi (c. 780–850 CE), a Tajik mathematician and astronomer from Khwarazm, synthesised Greek and Indian mathematics in his work "Kitab al-mukhtasar fi hisab al-jabr wal-muqabala", in which the principles of algebra were for the first time systematically expounded [16, 17]. As R. Rashed showed, the works of al-Khwarizmi represent a synthesis of Hellenistic and Indian traditions that penetrated the Islamic world in part through Sasanian mediation [16, 17]. J. Høyrup established that al-Khwarizmi's geometric justifications trace back to Babylonian and Greek traditions [53, 54, 55].

Abu Bakr Muhammad ibn Zakariyya al-Razi (865–925 CE) — the foremost Tajik physician and philosopher from Rayy — epitomises the generation of scholars who not merely absorbed the translated texts but creatively developed them [56, 57]. Appointed head of the Baghdad hospital, al-Razi compiled the medical encyclopaedia "Kitab al-Hawi" ("The Comprehensive Book"), which incorporated reworked translations of Galen, Indian medical observations, and his own clinical experience [57, 58, 59]. P. Pormann convincingly showed that al-Razi's clinical methodology grew directly out of the tradition connected with Gondishapur [60]. Moreover, al-Razi made significant contributions to alchemy and pharmacology [61], and in philosophy developed an original system combining elements of Platonism and the ethics of pleasure [56, 62, 63].

Abu Nasr Muhammad al-Farabi (c. 872–950 CE) — a native of Farab in Transoxiana and one of the foremost thinkers of medieval Islam — embodies the type of scholar who, drawing upon the translation heritage of Bayt alHikma, brought Arabic-language philosophy to a qualitatively new level [64]. Having arrived in Baghdad after the peak of the translation movement, al-Farabi found there a richest library of Arabic translations of Aristotle and Plato — the direct heritage of the Gondishapur tradition, assimilated through Bayt al-Hikma. M. Mahdi in his fundamental study shows that al-Farabi's political philosophy — primarily the concept of the "virtuous city" (Madinat al-fadila) — represents an original synthesis of Plato's "Republic" and Aristotle's "Politics", achieved for the first time in Arabic [65]. The critical edition of this work was prepared by R. Walzer [66]. In addition to political philosophy, al-Farabi developed the theory of logic as a universal science, expanding Aristotle's "Organon" to include rhetoric and poetics, as studied in detail in "The Cambridge Companion to Arabic Philosophy" [67]. T.-A. Druart notes that al-Farabi was the one who formed the canon of philosophical education subsequently used by Ibn Sina [68]. No less significant is his contribution to music theory: the work "Kitab al-musiqa al-kabir" ("The Great Book of Music") was the first systematic analysis of the theoretical foundations of music in Arabic, uniting the Pythagorean, Aristoxenian and Indian traditions; D.M. Randel convincingly showed that it was through this work that Arabic music theory entered medieval Latin scholarship [69].

Ahmad ibn Muhammad al-Farghani (Alfraganus, died after 857 CE) — an outstanding Tajik astronomer and mathematician from Fergana — became one of the first scholars at Bayt al-Hikma to systematise the Ptolemaic astronomical canon for a new audience [70, 71]. His principal work "Jawami' ʿilm al-nujum" ("The Compendium of Astronomy", written c. 833 CE) is a concise non-mathematical exposition of the "Almagest": it describes the movements of the celestial spheres, distances and sizes of the planets and stars, and the duration of seasons. B. Abdukhalimov in a dedicated study of this work shows that al-Farghani consistently adopted the parameters of the Ptolemaic geocentric system, correcting a number of numerical values on the basis of new observations conducted at al-Mamun's court [72]. R. Lorch published a critical edition of another important work by al-Farghani — a treatise on the astrolabe, recognised as the oldest surviving Arabic treatise on the construction and use of this instrument [73]. Exceptionally great is al-Farghani's influence on European medieval science: the "Compendium" was twice translated into Latin in the 12th century — by John of Seville and Herman of Carinthia — and until the 16th century remained in Europe the principal textbook of Arabic astronomy, serving as a source for Dante, Roger Bacon and Christopher Columbus [70].

The Banu Musa brothers (Muhammad, Ahmad and Hasan ibn Musa ibn Shakir, flourished c. 820–870 CE) — Persian scholars from Khorasan, educated at al-Mamun's court — played a key role both in mathematics and mechanics, and in organising the translation movement itself [15, 74]. Their "Book on the Measurement of Plane and Spherical Figures" (Kitab maʿrifat masahat al-ashkal) contains rigorous proofs of theorems on the measurement of a circle, on the surface area and volume of a sphere, and on the trisection of an angle; R. Rashed established that in proving certain theorems the brothers employed a genuinely "infinitesimal" approach anticipating the methods of Cavalieri [75]. Another outstanding contribution was the "Book of Ingenious Devices" (Kitab al-hiyal) — a collection of descriptions of nearly a hundred automatic mechanisms, including self-filling vessels, automatic fountains and musical instruments; D.R. Hill, who prepared the first complete English translation of this work, showed that it was the first in the history of engineering to apply the principles of automatic fluid flow regulation [76]. No less important is the brothers' role as patrons of translation: it was on their commission and with their funds that Thabit ibn Qurra executed first-rate translations of Archimedes, Apollonius and Menelaus; M. Clagett in the monumental edition "Archimedes in the Middle Ages" traces in detail how Latin translations of these Arabic versions determined the development of European mathematics in the 12th–14th centuries [77]. R. Rashed in a dedicated article demonstrates that it was the Banu Musa who ensured the assimilation and further development of Archimedean mathematics in the Arab world [78].

Abu Ali Yahya ibn Abi Mansur (died c. 830 CE) — court astronomer and director of al-Mamun's observational programme — is the central figure of the first systematic astronomical reform in the Islamic world [79, 80]. By order of the caliph, he headed the al-Shammassiyya observatory in Baghdad and the parallel observatory at Dayr Murran in Damascus, where a group of scholars conducted coordinated observations of the Sun, Moon and planets in order to verify the Ptolemaic parameters [81]. The result was the "Zij al-mumtahan" ("The Verified Astronomical Tables"), published in facsimile by F. Sezgin [82] and studied in numerous works by E.S. Kennedy [83]. The latter established that the al-Mumtahan tables contain systematically corrected — compared with Ptolemy — values of the parameters of solar and lunar motion, the result of genuine observations and not mere copying of ancient data. M. Viladrich showed that the planetary latitude tables in the zij use Indian (rather than Ptolemaic) values, attesting to the sustained influence of the Sasanian-Indian astronomical tradition [80]. The group of scholars under Yahya's leadership also conducted the famous measurement of a meridian arc on the Sinjar plain, confirming and refining Ptolemy's data on the length of a degree; D.A. King describes this experiment as the first geodetic measurement in Islamic science to aspire to independent empirical verification of ancient data [80, 81].

Abu Maʿshar al-Balkhi (787–886 CE) — a Tajik astrologer and astronomer from Balkh (Khorasan) — played an exceptional role in the synthesis of Sasanian, Indian and Greek astrological traditions and in the subsequent transmission of this synthesis to medieval Europe [84, 85]. His principal work "Kitab al-madkhal al-kabir" ("The Great Introduction to Astrology", written c. 848 CE) — the most complete systematic exposition of astrology in Arabic — was twice translated into Latin in the 12th century (by John of Seville and Herman of Carinthia) and became, in the assessment of R. Lemay, the most important channel for the penetration of Aristotelian natural philosophy into Western European scholarship prior to the appearance of direct translations of Aristotle [1, 41]. The critical edition of the Arabic text with English translation was prepared by K. Yamamoto and C. Burnett [84]. Another significant work — "Kitab al-qiranat" ("The Book of Great Conjunctions") — develops the theory of historical astrology: Abu Maʿshar connected the succession of kingdoms, prophets and religions with the cycles of great conjunctions of Jupiter and Saturn, continuing the Sasanian historical-astrological tradition; an edition of this text was also carried out by K.

Yamamoto and C. Burnett [86]. D. Pingree in the monograph "The Thousands of Abu Maʿshar" reconstructed the astrologer's lost work (Kitab al-uluf) and demonstrated that it drew on Sasanian "millennial" astrological tables tracing back to the Indian-Persian synthesis that had taken shape as early as the era of Gondishapur [87]. P. Adamson showed that in "The Great Introduction" Abu Maʿshar constructs a genuine philosophical defence of astrology, drawing on Aristotle's theory of causality and al-Kindi's doctrine of the radiant action of celestial bodies [88], which makes him not merely a practical astrologer but a participant in the philosophical discussions of Bayt al-Hikma.

6. Medical Education and the Bimaristan System

One of the most tangible manifestations of Gondishapur's influence in the Islamic world was the system of hospitals (bimaristans), reproduced according to the Persian model. The very word "bimaristan" is of Persian origin (bimar + stan, "place for the sick"). A. Miller demonstrated that the Gondishapur hospital was the first institution in history where systematic clinical training was conducted directly at the bedside [24]. M. Dols, however, points to the need to take into account other sources of the Islamic hospital tradition — including late antique and Byzantine ones [18].

According to the sources, the first large bimaristan in Baghdad ("al-Rashidi") was founded on the model of the Gondishapur one: with a permanent staff of physicians of various specialisations, twenty-four-hour patient care, and a system of clinical supervisors [33, 29]. Subsequently this model spread throughout the caliphate [89, 90]. G. Leiser in his study of medical education in the Islamic world from the 7th to the 14th centuries shows that clinical lectures, ward rounds with students, and examination attestation, as practised in Gondishapur, were reproduced in Baghdad [25]. A. Ragab in the monograph on the medieval Islamic hospital traces this continuity in detail [33].

M. Hossein Azizi in his article in "Archives of Iranian Medicine" states that Gondishapur was "the most important medical centre of Antiquity", and considers its hospital as the prototype of Islamic bimaristans [20]. This view is shared by H.D. Modanlou, who describes Gondishapur as "the world's first academic medicine" [22].

DISCUSSION

A comparative structural analysis of the Academy of Gondishapur and Bayt al-Hikma reveals a number of fundamental parallels. Both institutions functioned under the patronage of a ruler and were administered by a prominent scholar closely connected with the sovereign. Both had extensive book collections and encouraged the attraction of talent regardless of confessional and ethnic background. Medicine occupied a central place in both institutions — in Gondishapur directly through the hospital, and in Bayt al-Hikma through its close connection with the bimaristans [1, 12, 19].

At the same time, significant differences are also traceable. D. Gutas convincingly shows that Bayt al-Hikma specialised primarily in book storage and translation activity, whereas Gondishapur placed emphasis on teaching and the application of knowledge [1]. Gutas himself critically revises the traditional narrative of a "direct line" from Gondishapur to Bayt al-Hikma, pointing out that the translation movement had more complex political and social determinants [1, 91]. Nevertheless, the personal connections and methodological continuity are beyond doubt: Hunayn ibn Ishaq — a student of a Gondishapur graduate — became the central figure of the translation project [43, 46].

Sabra's concept of "appropriation and naturalisation" makes it possible to reinterpret the mechanism of continuity: what is involved is not a passive reproduction of Gondishapur models, but a creative adaptation in a new political and intellectual context [11]. S. Montgomery examines this process in the categories of the "mobility of science": knowledge moves together with people, texts and institutional models [42]. G. Saliba, in turn, accentuates the independent contribution of Islamic science, not reducible to mere reception [14].

The connection of Gondishapur with Nestorian Christian scholarship, analysed by A.H. Becker [23] and S. Brock [39], explains the bilingualism (Greek/Syriac) of the translation tradition inherited by Bayt al-Hikma. J. Watt shows that it was precisely Syriac-speaking scholars who in the early Abbasid period served as the chief intermediaries between Greek philosophy and the Arabic-speaking audience [40]. This makes the Nestorian component of Gondishapur fundamentally important for understanding the Abbasid translation movement.

In the field of mathematics and astronomy, Iranian mediation is revealed in the fact that Indian numeral systems and astronomical tables reached Baghdad through the filter of Sasanian science. D. Pingree established that the "Zij al-Shahriyar" — astronomical tables compiled in Sasanian Iran — formed the basis of the first Arabic astronomical works [26, 27]. J. Høyrup demonstrated that the formation of "Islamic mathematics" was a process of synthesising several traditions, with the Perso-Sasanian channel serving as one of the main conduits [55].

## CONCLUSION

The study carried out makes it possible to formulate the following conclusions. First, the Academy of Gondishapur served as the principal institutional precursor of Bayt al-Hikma: Gondishapur graduates became the first physicians and scholars of the Abbasid court, transferring to Baghdad both specific specialists and the organisational model of clinical training and hospital practice. Second, the translation movement of the 8th–10th centuries drew substantially on the translation infrastructure and methodology that had formed at Gondishapur during the Sasanian period, although — as critical studies show [1, 18, 11] — this involved a complex, multi-level process and not a mechanical "transfer".

Third, the synthesis of Greek, Persian and Indian knowledge accomplished in Bayt al-Hikma was prepared precisely by the multicultural and multilingual character of Gondishapur. Finally, fourth, the most significant figures of the science of the "golden age" — from Hunayn ibn Ishaq to al-Razi and al-Khwarizmi — either directly represented the Gondishapur tradition or were taught by those who had emerged from it. All this allows one to assert that without the institutional and personnel contribution of Gondishapur, the Islamic "golden age" of science could scarcely have taken place in so brief a historical time span [28, 2, 4, 14].

## AI USE STATEMENT

This work utilized artificial intelligence tools (ChatGPT 5.4 by OpenAI and Claude Sonnet 4.6 by Anthropic) to support literature exploration, text editing, and English language translation. These tools were used solely for assistance, and all content has been critically reviewed and validated by the authors. The authors retain full responsibility for the accuracy and integrity of the work.